\documentclass[review,12pt,authoryear]{elsarticle}

\usepackage{multirow}
\usepackage{amsmath,xfrac,cases,mathrsfs} 
\numberwithin{equation}{section}
\usepackage{amssymb}
\usepackage{enumerate}
\usepackage{lineno,hyperref}
\usepackage{braket}

\newtheorem{thm}{Theorem}
\newtheorem{lem}[thm]{Lemma}
\newdefinition{rmk}[thm]{Remark}
\newdefinition{prop}[thm]{Proposition}
\newproof{pf}{Proof}

\journal{Journal of Statistical Planning and Inference}

\begin{document}
	
\begin{frontmatter}
		
\title{Least squares estimators for reflected Ornstein–Uhlenbeck processes}

\author[rvt]{Han Yuecai}
\ead{hanyc@jlu.edu.cn}
		
\author[rvt]{Zhang Dingwen\corref{cor1}}		\ead{zhangdw20@mails.jlu.edu.cn}
		
\cortext[cor1]{Corresponding author}

\address[rvt]{School of Mathematics, Jilin University, Changchun, China}
	
\begin{abstract}
In this paper, we investigate the parameter estimation problem for reflected Ornstein–Uhlenbeck processes. Both the estimates based on continuously observed processes and discretely observed processes are considered. The explicit formulas for the estimators are derived using the least squares method. Under some regular conditions, we obtain the consistency and establish the asymptotic normality for the estimators. Numerical results show that the proposed estimators perform well with moderate sample sizes.
\end{abstract}
		
		
\end{frontmatter}
	
	
\section{Introduction}
Consider a filtered probability space $(\Omega, \mathcal{F}, \{\mathcal{F}_{t}\}_{t\geq 0}, \mathbb{P})$ where the filtration $\{\mathcal{F}_{t}\}_{t\geq 0}$ satisfies the usual conditions. Let $W=\{W_{t}\}_{t\geq 0}$ be a standard Brownian motion adapted to $\{\mathcal{F}_{t}\}_{t\geq 0}$. The reflected Ornstein–Uhlenbeck (OU) process reflected at $0$ is described by the following stochastic differential equation (SDE)
\begin{equation}\label{eq1}
\left\{
\begin{aligned}
&\mathrm{d}X_{t}=-\theta X_{t}\mathrm{d}t+\sigma\mathrm{d}W_{t}+\mathrm{d}L_{t},\\
&X_{t}\geq 0 \quad\text{for all}\quad t\geq0,\\
&X_{0}=x,
\end{aligned}
\right.
\end{equation}
where $\theta\in (0,\infty)$ is the unknown parameter, $\sigma\in (0,\infty)$ is a constant and $L=\{L_{t}\}_{t\geq 0}$ is the minimal continuous increasing process which ensures that $X_{t}\geq 0$ for all $t\geq 0$. The process $L$ increases only when $X$ hits the boundary $0$, so that
\begin{equation*}
\int_{[0, \infty)} I(X_{t}\geq 0) \mathrm{d} L_{t}=0,
\end{equation*}
where $I(\cdot)$ is the indicator function. 

The reflected OU process behaves like a standard OU process in the interior of its domain $(0, \infty)$. Benefiting from its reflecting barrier, the reflected OU process has been widely used in many areas such as the queueing system \citep{Ward2005}, financial engineering \citep{Bo2010} and mathematical biology \citep{Ricciardi1987}. The reflecting barrier is assumed to be $0$ for the physical restriction of the state processes such as queue-length, stock prices and interest rates, which take non-negative values. For more details on reflected OU processes and their broad applications, one can refer to \cite{Harrison1985} and \cite{Whitt2002}.
	
The parameter estimation problem in the reflected OU process has gained much attention in recent years due to its increased applications in broad fields. It is necessary that the parameters which characterize the reflected OU process should be estimated via the data in many real-world applications. 	
As far as we know, the maximum likelihood estimator (MLE) for the drift parameter $\theta$ is studied in \cite{Bo2011}. They obtain the strong consistency and asymptotic normality of their estimator, but they don't get the explicit form of asymptotic variance. The sequential MLE based on the continuously observed processes throughout a random time interval $[0,\tau]$ is studied in \cite{Lee2012}, where $\tau$ is a stopping time. The main tool used in the above two papers is the Girsanov's theorem of reflected Brownian motion.
On the other hand, an ergodic type of estimator for $\theta$ based on discrete observations is studied in \cite{Hu2015}. Recently, the moment estimators for all parameters $(\theta,\sigma)$ based on the ergodic theorem is studied in \cite{Hu2021}. However, there is only limited literature on least squares estimator (LSE) for the drift parameter of a reflected OU process.

In this paper, we propose two types of LSEs for the drift parameter $\theta$ based on continuously observed processes and discretely observed processes respectively. 
The continuous-type LSE is motivated by aiming to minimize
\begin{equation*}
\int_{0}^{T}\left|\dot{X}_{t}+\theta X_{t}-\dot{L}_{t}\right|^{2} \mathrm{d}t.
\end{equation*}
It is a quadratic function of $\theta$, although we don't know $\dot{L}_{t}$ and $\dot{X}_{t}$. The minimum is achieved when 
\begin{equation*}
\hat{\theta}_{T}=-\frac{\int_{0}^{T} X_{t}\mathrm{d} X_{t}-\int_{0}^{T}X_{t}\mathrm{d}L_{t}}{\int_{0}^{T} X_{t}^{2} \mathrm{d} t}.
\end{equation*}
Assume that $h\rightarrow0$ and $nh\rightarrow\infty$, as $n\rightarrow\infty$. When the processes is observed at the discrete time instants $\{t_{k}=kh, k=0,1,\cdots,n\}$, the discrete-type LSE is motivated by minimizing the following contrast function
\begin{equation*}
\sum_{k=0}^{n-1}|X_{t_{k+1}}-X_{t_{k}}+\theta X_{t_{k}}h-\vartriangle_{k}L|^{2},
\end{equation*}
where $\vartriangle_{k}L=L_{t_{k+1}}-L_{t_{k}}$. The minimum is achieved when
\begin{equation*}
\tilde{\theta}_{n}=-\frac{\sum_{k=0}^{n-1}X_{t_{k}}(X_{t_{k+1}}-X_{t_{k}}-\vartriangle_{k}L)}{\sum_{k=0}^{n-1}X_{t_{k}}^{2}h}.
\end{equation*}
The remainder of this paper is organized as follows. In Section \ref{sec2}, we describe some preliminary results related to our context. Section \ref{sec3}  is devoted to obtaining the asymptotic behavior of the two estimators. Section \ref{sec4} presents some numerical results and Section \ref{sec5} concludes. 

\section{Preliminaries}\label{sec2}
In this section, we first introduce some basic facts. Throughout this paper, we shall use the notation ``$\stackrel{P}{\longrightarrow}$" to denote ``convergence in probability" and the notation ``$\sim$" to denote ``convergence in distribution".

With the previous results \citep{Hu2015,Linetsky2005,Ward2003}, we know that the unique invariant density of $\{X_{t}\}_{t\geq 0}$ is
\begin{equation}\label{eq2}
p(x)=2 \sqrt{\frac{2 \gamma}{\sigma^{2}}} \phi\left(\sqrt{\frac{2 \gamma}{\sigma^{2}}} x\right), \quad x \in[0, \infty),
\end{equation}
where $\phi(u)=(2 \pi)^{-1 / 2} e^{-\frac{u^{2}}{2}}$ is the (standard) Gaussian density function. 

Based on the basic stability theories of Markov processes, we have the following ergodic lemma.

\begin{lem}\label{lem1}
For any $x \in \mathbb{R}_{+}$ and any $f \in L_{1}(\mathbb{R}_{+}, \mathcal{B}(\mathbb{R}_{+}))$, we have
\begin{enumerate}[a.]
	\item The continuously observed processes $\{X_{t}\}_{t\geq0}$ is ergodic,
	\begin{equation*} 
	\lim _{T \rightarrow \infty} \frac{1}{T} \int_{0}^{T}f(X_{t})\mathrm{d}t=\mathbb{E}[f(X_{\infty})]=\int_{0}^{\infty} f(x) p(x) d x.
	\end{equation*}
	\item The discretely observed processes $\{X_{t_{k}}, k=0,1\cdots,n\}$ is ergodic,
	\begin{equation*}
	\lim _{n \rightarrow \infty} \frac{1}{n} \sum_{k=1}^{n} f(X_{t_{k}})=\mathbb{E}[f(X_{\infty})]=\int_{0}^{\infty} f(x) p(x) d x.
	\end{equation*}
\end{enumerate}
\end{lem}
\noindent\textbf{Proof of Lemma \ref{lem1}}. One can see  \cite{Han2016} and \cite{Hu2015} for a proof.
\hfill$\square$

By Lemma \ref{lem1} and the unique invariant density Eq. (\ref{eq2}), we obtain the formula of second-order moment estimator as following
\begin{equation}\label{eq3}
\lim_{T \rightarrow \infty}\frac{1}{T}\int_{0}^{T}X^{2}_{t}\mathrm{d}t=\lim_{n\rightarrow \infty}\frac{1}{n}\sum_{k=1}^{n}X_{t_{k}}^{2}=\mathbb{E}|X(\infty)|^{2}=\int_{0}^{\infty} x^{2} p(x) \mathrm{d} x=\frac{\sigma^{2}}{2\theta}.
\end{equation}

\section{Asymptotic behavior of the least squares estimators}\label{sec3}
In this section, we consider the asymptotic behavior of the LSEs for the drift parameter $\theta$. By Eq. (\ref{eq1}), we provide two useful and crucial alternative expressions for $\hat{\theta}_{T}$ and $\tilde{\theta}_{n}$
\begin{equation}\label{eq4}
\hat{\theta}_{T}=\theta-\sigma \frac{\int_{0}^{T} X_{t} \mathrm{d} W_{t}}{\int_{0}^{T} X_{t}^{2} \mathrm{d} t},
\end{equation}
and 
\begin{equation}\label{eq5}
\tilde{\theta}_{n}=\theta+\frac{\sum_{k=0}^{n-1}X_{t_{k}}\int_{t_{k}}^{t_{k+1}}\theta(X_{t}-X_{t_{k}})\mathrm{d}t-\sigma\vartriangle_{k}W}{\sum_{k=0}^{n-1}X^{2}_{t_{k}}h}.
\end{equation}

The following theorem proves the consistency of the continuous-type LSE.
\begin{thm}\label{thm1}
The continuous-type LSE $\hat{\theta}_{T}$ of $\theta$ admits the strong consistency, i.e.,
\begin{equation*}
\hat{\theta}_{T}\stackrel{P}{\longrightarrow} \theta,
\end{equation*}
as $T$ tends to infinity.
\end{thm}
\noindent\textbf{Proof of Theorem \ref{thm1}}. From the alternative expression Eq. (\ref{eq4}), we have
\begin{equation*}
\hat{\theta}_{T}-\theta=-\sigma \frac{\frac{1}{T}\int_{0}^{T} X_{t} \mathrm{d} W_{t}}{\frac{1}{T}\int_{0}^{T} X_{t}^{2} \mathrm{d} t}.
\end{equation*}
By Lemma \ref{lem1} and Eq. (\ref{eq3}), we have 
\begin{equation}\label{eq6}
\lim_{T \rightarrow \infty}\frac{1}{T}\int_{0}^{T}X_{t}^{2}\mathrm{d}t= \frac{\sigma^{2}}{2\theta}.
\end{equation}
Taking into account that the process $\{\int_{0}^{t}X_{s}\mathrm{d}W_{s}, t\geq 0\}$ is a martingale and with quadratic variation $\int_{0}^{t}X_{s}^{2}\mathrm{d}s$. Then 
\begin{equation*}
\mathbb{E}\bigg[\frac{1}{T}\int_{0}^{T}X_{t}\mathrm{d}W_{t}\bigg]=0,
\end{equation*}
and 
\begin{equation*}
\lim_{T \rightarrow \infty}\mathbb{E}\bigg[\big(\frac{1}{T}\int_{0}^{T}X_{t}\mathrm{d}W_{t}\big)^{2}\bigg]=\frac{\sigma^{2}}{2\theta T}=O(T^{-1}).
\end{equation*}
By Chebyshev's inequality, we have
\begin{equation}\label{eq7}
	\lim_{T\rightarrow\infty}\frac{1}{T}\int_{0}^{T}X_{t}\mathrm{d}W_{t}=0.
\end{equation}
Combining Eq. (\ref{eq6}) and (\ref{eq7}), we obtain the desired results.
\hfill$\square$

We establish the asymptotic normality of the continuous-type LSE in the following theorem. The convergence rate is comparable to MLE based approch \citep{Bo2011}, and we obtain the explicit formula of the asymptotic variance. 
\begin{thm}\label{thm2}
	The continuous-type LSE $\hat{\theta}_{T}$ of $\theta$ admits the asymptotic normality, i.e.,
	\begin{equation*}
		\sqrt{T}(\hat{\theta}_{T}-\theta)\sim\mathcal{N}(0, 2\theta),
	\end{equation*}
	as $T$ tends to infinity.
\end{thm}
\noindent\textbf{Proof of Theorem \ref{thm2}}
Note that
\begin{equation*}
	\begin{aligned}
		\sqrt{T}(\hat{\theta}_{T}-\theta)&=-\sigma\sqrt{T}\frac{\int_{0}^{T}X_{t}\mathrm{d}W_{t}}{\int_{0}^{T} X_{t}^{2}\mathrm{d}t}\\
		&=-\frac{\frac{\sigma}{\sqrt{T}}\int_{0}^{T}X_{t}\mathrm{d}W_{t}}{\frac{1}{T}\int_{0}^{T} X_{t}^{2}\mathrm{d}t}.
	\end{aligned}
\end{equation*}
From Eq. (\ref{eq3}), we have that $\frac{1}{T}\int_{0}^{T} X_{t}^{2}\mathrm{d}t$ converges to $\frac{\sigma^{2}}{2\theta}$
almost surely, as $T$ tends to infinity. Then, it is sufficient to show that $\frac{\sigma}{\sqrt{T}}\int_{0}^{T}X_{t}\mathrm{d}W_{t}$ converges in law to a centered normal distribution as $T$ tends to infinity. It follows immediately from $X_{t}$ is adapted with respect to $\mathcal{F}_{t}$ that $\frac{\sigma}{\sqrt{T}}\int_{0}^{T}X_{t}\mathrm{d}W_{t}$ is a Gaussian random variable with mean $0$ and variance $\frac{\sigma^{2}}{T}\int_{0}^{T} X_{t}^{2}\mathrm{d}t$. Based on Eq. (\ref{eq3}) again, we obtain
\begin{equation}\label{eq8}
	\frac{\sigma}{\sqrt{T}}\int_{0}^{T}X_{t}\mathrm{d}W_{t}\sim\mathcal{N}(0, \frac{\sigma^{4}}{2\theta}).
\end{equation}
By Slutsky's theorem and Eq. (\ref{eq8}), we have
\begin{equation*}
	\frac{\frac{\sigma}{\sqrt{T}}\int_{0}^{T}X_{t}\mathrm{d}W_{t}}{\frac{1}{T}\int_{0}^{T} X_{t}^{2}\mathrm{d}t}\sim\mathcal{N}(0,2\theta),
\end{equation*}
which completes the proof.
\hfill$\square$

The following theorem proves the consistency of the discrete-type LSE.
\begin{thm}\label{thm3}
	The discrete-type LSE $\tilde{\theta}_{n}$ admits the consistency, i.e.,
	\begin{equation*}
		\tilde{\theta}_{n}\stackrel{P}{\longrightarrow}\theta,
	\end{equation*}
	as $n$ tends to infinity.
\end{thm}
\noindent\textbf{Proof of Theorem \ref{thm3}}. From the alternative expression Eq. (\ref{eq5}), we have
\begin{equation*}
\tilde{\theta}_{n}-\theta=\frac{\frac{1}{nh}\sum_{k=0}^{n-1}X_{t_{k}}\big(\int_{t_{k}}^{t_{k+1}}\theta(X_{t}-X_{t_{k}})\mathrm{d}t-\sigma\vartriangle_{k}W\big)}{\frac{1}{nh}\sum_{k=0}^{n-1}X^{2}_{t_{k}}h}.
\end{equation*}
We first consider the estimate of $\sup_{t_{k}\leq t\leq t_{k+1}}|X_{t}-X_{t_{k}}|$. For $t_{k}\leq t\leq t_{k+1}$, we have
\begin{equation*}
\begin{aligned}
&|X_{t}-X_{t_{k}}|\\
=&|-\theta\int_{t_{k}}^{t} X_{u}\mathrm{d}u+\sigma(W_{t}-W_{t_{k}})+(L_{t}-L_{t_{k}})|\\
=&|-\theta\int_{t_{k}}^{t}(X_{u}-X_{t_{k}})\mathrm{d}u-\theta X_{t_{k}}(t-t_{k})+\sigma(W_{t}-W_{t_{k}})(L_{t}-L_{t_{k}})|\\
\leq&|X_{t_{k}}|h\theta+\sup_{t_{k}\leq t\leq t_{k+1}}\big(\sigma|W_{t}-W_{t_{k}}|+(L_{t}-L_{t_{k}})\big)+\theta\int_{t_{k}}^{t}|X_{u}-X_{t_{k}}|\mathrm{d}u.
\end{aligned}
\end{equation*}
By Gronwall's inequality, we have
\begin{equation*}
|X_{t}-X_{t_{k}}|\leq\bigg(|X_{t_{k}}|h\theta+\sup_{t_{k}\leq t\leq t_{k+1}}\big(\sigma|W_{t}-W_{t_{k}}|+(L_{t}-L_{t_{k}})\big)\bigg)e^{\theta(t-t_{k})}.
\end{equation*}
It follows that
\begin{equation*}
\sup_{t_{k}\leq t\leq t_{k+1}}|X_{t}-X_{t_{k}}|\leq \bigg(|X_{t_{k}}|h+\sup_{t_{k}\leq t\leq t_{k+1}}\big(\sigma|W_{t}-W_{t_{k}}|+(L_{t}-L_{t_{k}})\big)\bigg)e^{\theta h}.
\end{equation*}
By the properties of the process $L$, we have
\begin{equation*}
L_{t_{k+1}}-L_{t_{k}}=\max\big(0, A_{t_{k}}-X_{t_{k}}\big),
\end{equation*}
where $A_{t_{k}}=\sup_{t_{k}\leq t\leq t_{k+1}}\big\{\theta X_{t_{k}}(t-t_{k})-\sigma(W_{t}-W_{t_{k}})\big\}$.
By all paths of Brownian motion are $\alpha$-H$\ddot{o}$lder continuity, where $\alpha\in(0,\frac{1}{2})$, we have
\begin{equation*}
\begin{aligned}
\sup_{t_{k}\leq t\leq t_{k+1}}|X_{t}-X_{t_{k}}|&\leq Ch^{\alpha}e^{\theta h}=O(h^{\alpha}),
\end{aligned}
\end{equation*}
where $C$ is a constant.
Then 
\begin{equation}\label{eq10}
\begin{aligned}
&\frac{1}{nh}\sum_{k=0}^{n-1}X_{t_{k}}\int_{t_{k}}^{t_{k+1}}\theta(X_{t}-X_{t_{k}})\mathrm{d}t\\
\leq&\frac{\theta}{nh}\sum_{k=0}^{n-1}X_{t_{k}}\sup_{t_{k}\leq t\leq t_{k+1}}|X_{t}-X_{t_{k}}|h\\
=&O(h^{\alpha}),
\end{aligned}
\end{equation}
which goes to $0$ as $h\rightarrow0$.
 
Let $\phi_{k}(t)=X_{t_{k}}I_{\{t\in[t_{k},t_{k+1})\}}(t)$. Then we have
\begin{equation}\label{eq11}
\lim_{n \rightarrow \infty}\sum_{k=0}^{n-1}X_{t_{k}}\vartriangle_{k}W=\lim_{n \rightarrow \infty}\sum_{k=0}^{n-1}\int_{0}^{nh}\phi_{k}(t)\mathrm{d}W_{t}.
\end{equation}
By some similar arguments as in the proof of Theorem \ref{thm1}, we have 
\begin{equation}\label{eq12}
\lim_{n\rightarrow\infty}\frac{1}{nh}\sum_{k=0}^{n-1}X_{t_{k}}\sigma\vartriangle_{k}W=0.
\end{equation}
Combining Eq. (\ref{eq3}), (\ref{eq10}) and (\ref{eq12}), we obtain the desired results.
\hfill$\square$\\

The following theorem establishes the asymptotic normality of the discrete-type LSE.

\begin{thm}\label{thm4}
Assume that $nh^{1+2\alpha}\rightarrow 0$ for $\alpha\in(0,1/2)$, as $n$ tends to infinity. The discrete-type LSE $\tilde{\theta}_{n}$ of $\theta$ admits the asymptotic normality, i.e.,
\begin{equation*}
\sqrt{nh}(\tilde{\theta}_{n}-\theta)\sim\mathcal{N}(0,2\theta),
\end{equation*}
as $n$ tends to infinity.
\end{thm}
\noindent\textbf{Proof of Theorem \ref{thm4}}. Note that
\begin{equation*}
\sqrt{nh}(\tilde{\theta}_{n}-\theta)=\frac{\frac{1}{\sqrt{nh}}\sum_{k=0}^{n-1}X_{t_{k}}\big(\int_{t_{k}}^{t_{k+1}}\theta(X_{t}-X_{t_{k}})\mathrm{d}t-\sigma\vartriangle_{k}W\big)}{\frac{1}{nh}\sum_{k=0}^{n-1}X^{2}_{t_{k}}h}.
\end{equation*}
By Eq. (\ref{eq10}), we have
\begin{equation}
\frac{1}{\sqrt{nh}}\sum_{k=0}^{n-1}X_{t_{k}}\int_{t_{k}}^{t_{k+1}}\theta (X_{t}-X_{t_{k}})\mathrm{d}t\leq O(\sqrt{nh^{1+2\alpha}}),
\end{equation}
which goes to $0$ as $n$ tends to infinity. By some similar arguments as in the proof of Theorem \ref{thm2}, we have 
\begin{equation*}
\frac{1}{nh}\int_{0}^{nh}\phi_{k}(t)\mathrm{d}W_{t}\sim\mathcal{N}(0,\frac{\sigma^{4}}{2\theta}).
\end{equation*}
By Eq. (\ref{eq11}), we have
\begin{equation*}
\frac{\sigma}{\sqrt{nh}}\sum_{k=0}^{n-1}X_{t_{k}}\vartriangle_{k}W\sim\mathcal{N}(0,\frac{\sigma^{4}}{2\theta}).
\end{equation*}
By Eq. (\ref{eq3}) and Slutsky's theorem, we obtain the desired results. 
\hfill$\square$

\begin{rmk}
Our method can be applied to the reflected OU processes with two-sided reflecting barriers $(0,b)$, where $b\in(0,\infty)$. The two types of LSEs of a two-sided reflected OU process are 
\begin{equation*}
\hat{\theta}_{T}=-\frac{\int_{0}^{T} X_{t}\mathrm{d} X_{t}-\int_{0}^{T}X_{t}\mathrm{d}L_{t}+\int_{0}^{T}X_{t}\mathrm{d}R_{t}}{\int_{0}^{T} X_{t}^{2} \mathrm{d} t},
\end{equation*} 
and 
\begin{equation*}
\tilde{\theta}_{n}=-\frac{\sum_{k=0}^{n-1}X_{t_{k}}(X_{t_{k+1}}-X_{t_{k}}-\vartriangle_{k}L+\vartriangle_{k}R)}{\sum_{k=0}^{n-1}X_{t_{k}}^{2}h},
\end{equation*}
\end{rmk}
where $R$ is the minimal continuous increasing process such that $X\leq b$. The unique invariant density is given by \citep{Linetsky2005}
\begin{equation*}
p(x)=\frac{\sqrt{2 \theta}}{\sigma} \frac{\phi\left(\frac{\sqrt{2 \theta}}{\theta} x\right)}{\Phi\left(\frac{\sqrt{2 \theta}}{\sigma} b\right)-\frac{1}{2}}, \quad x \in[0, b],
\end{equation*}
where $\Phi(y)=\int_{-\infty}^{y}\phi(u)\mathrm{d}u$. 
Hence the consistency and asymptotic distributions of the two estimators could be obtained. The proofs are similar to the proofs of Theorem \ref{thm1}-\ref{thm4}. We omit the details here.

\section{Numerical results}\label{sec4}
In this section, we present some numerical results. For a Monte Carlo simulation of the reflected OU process, one can refer to \cite{Lepingle(1995)}, which is known to yield the same rate of convergence as the usual Euler–Maruyama scheme. 

Denote the time between each discretization step by $h=0.01$. We perform $N=1000$ Monte Carlo simulations of the sample paths generated by the model with different settings. The overall parameter estimates are evaluated by the bias, standard deviation (Std.dev) and mean squared error (MSE). We also give calculations for the asymptotic variance (Asy.var) $\sqrt{nh}(\tilde{\theta}_{n}-\theta)$. The results are presented in Table \ref{table1}.

What we need to emphasize is that the asymptotic variance is exactly the approximation of $2\theta$ even with different settings of $\theta$. It is effective to verify the explicit, closed form formula proposed in Theorem \ref{thm2} and \ref{thm4}.

Table \ref{table1} summarizes the main findings over 1000 simulations. We observe that as the sample size increases, the bias decreases and is small, that the empirical and model-based standard errors agree reasonably well. The performance improves with larger sample
sizes.

\begin{figure}[h]
	\centering
	\includegraphics[scale=0.7]{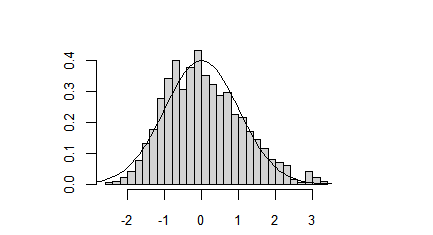} 
	\caption{Histogram of $\sqrt{T}(\hat{\theta}_{T}-\theta)$ with $T=100$, $h=0.01$, $\theta=0.5$ and $\sigma=0.2$.}
	\label{fig1}
\end{figure}

\begin{figure}[h]
	\centering
    \includegraphics[scale=0.7]{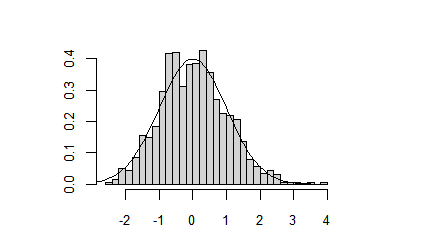} 
	\caption{Histogram of $\sqrt{T}(\hat{\theta}_{T}-\theta)$ with $T=1000$, $h=0.01$, $\theta=0.2$ and $\sigma=0.2$.}
	\label{fig2}
\end{figure}

The distribution of the proposed estimator with two different settings are illustrated as a histogram in Figure \ref{fig1} and \ref{fig2}. In each figure, the standard normal distribution density is overlayed as a solid curve. The histogram asymptotically approximates to the standard normal distribution density. Thus, the LSEs work well whether $\theta$ is big ($\theta=1$) or small ($\theta=0.5$) and whether in a fairly short time ($T=10$) or long ($T=1000$) time.
\begin{table}[htp]
\caption{Simulation results}
\begin{tabular}{ccccc}
\hline 
\makebox[0.2\textwidth]{True parameter} & \makebox[0.1\textwidth]{} & \makebox[0.18\textwidth]{$n=10^{3}$} & \makebox[0.18\textwidth]{$n=10^{4}$} & \makebox[0.18\textwidth]{$n=10^{5}$}\tabularnewline
\hline 
\hline 
$\theta=0.5$, 
& Bias    & 0.2006 & 0.0129  & -0.0076\tabularnewline
$\sigma=0.2$ 
& Std.dev & 0.4380 & 0.1040  & 0.0310\tabularnewline
& Asy.var & 1.9100 & 1.0800  & 0.9620\tabularnewline
& MSE     & 0.2320 & 0.0110  & 0.0010\tabularnewline
\hline 
$\theta=0.5$, 
& Bias    & 0.1890 & 0.0171 & -0.0084\tabularnewline
$\sigma=0.5$
& Std.dev & 0.4550  & 0.1080  & 0.0315\tabularnewline
& Asy.var & 2.0700  & 1.1700  & 0.9900\tabularnewline
& MSE     & 0.2430  & 0.0120  & 0.0011\tabularnewline
\hline 
$\theta=1$, 
& Bias    & 0.1410 & -0.0124 & -0.0255\tabularnewline
$\sigma=1$
& Std.dev & 0.5030  & 0.1450  & 0.0439\tabularnewline
& Asy.var & 2.5300  & 2.1200  & 1.9300\tabularnewline
& MSE     & 0.2730  & 0.0213  & 0.0026\tabularnewline
\hline 
\end{tabular}
\label{table1}
\end{table}


\section{Conclusion}\label{sec5}
In this paper, we present two types of least squares estimators for the reflected Ornstein–Uhlenbeck process based on continuously observed processes and discretely observed processes respectively. The consistency and the asymptotic normality have been studied. Moreover, we derive the explicit formula of the asymptotic variance, which is $2\theta$. Numerical results show that the least squares estimators work well with different settings.
Some further research may include investigating the statistical inference for the other reflected diffusions.

\end{document}